# Bandgap narrowing in Mn doped GaAs probed by room-temperature photoluminescence

S. Prucnal[1,*], K. Gao[1], I. Skorupa[1], L. Rebohle[1], L. Vines[4], H. Schmidt[3], M. Khalid[1], Y. Wang[1], E. Weschke[5], W. Skorupa[1], J. Grenzer[1], R. Hübner[1], M. Helm[1,2], S. Zhou[1]

[1] *Helmholtz-Zentrum Dresden-Rossendorf, Institute of Ion Beam Physics and Materials Research, Bautzner Landstr. 400, 01328 Dresden, Germany*

[2] *Center for Advancing Electronics Dresden (cfaed), Technische Universität Dresden, 01062 Dresden, Germany*

[3] *Tech Univ Chemnitz, Dept Mat Nanoelect, Fac Elect Engn & Informat Technol, D-09126 Chemnitz, Germany*

[4] *Department of Physics/Centre for Materials Science and Nanotechnology, University of Oslo, P.O. Box 1048 Blindern, N-0316 Oslo, Norway*

[5] *Helmholtz-Zentrum Berlin für Materialien und Energie, Wilhelm-Conrad-Röntgen-Campus BESSY II, D-12489 Berlin, Germany*

Abstract

The electronic band structure of the (Ga,Mn)As system has been one of the most intriguing problems in solid state physics over the past two decades. Determination of the band structure evolution with increasing Mn concentration is a key issue to understand the origin of ferromagnetism. Here we present room temperature photoluminescence and ellipsometry measurements of $Ga_{100\%-x}Mn_xAs$ alloy. The up-shift of the valence-band is proven by the red shift of the room temperature near band gap emission from the $Ga_{100\%-x}Mn_xAs$ alloy with increasing Mn content. It is shown that even a doping by 0.02 at.% of Mn affects the valence-band edge and it merges with the impurity band for a Mn concentration as low as 0.6 at.%. Both X-ray diffraction pattern and high resolution cross-sectional TEM images confirmed full recrystallization of the implanted layer and GaMnAs alloy formation.

*Corresponding author: s.prucnal@hzdr.de

# 1. Introduction

Manganese (Mn) as a p-type dopant in the semiconductor GaAs has been known for more than forty years [1-8] and the first ferromagnetic (Ga,Mn)As alloy was fabricated by Ohno *et al.* three decades later [2]. Since then, (Ga,Mn)As has been the prototype of diluted magnetic semiconductors (DMS) being intensively investigated up to the present day. However, the electronic band structure and the nature of carriers mediating the ferromagnetism in (Ga,Mn)As are still controversial [9-12]. The Mn ions substituting Ga in (Ga,Mn)As act as acceptors providing holes to mediate ferromagnetic interactions between the $S$=5/2 moments of the half-filled 3$d$ Mn shells. The single Mn acceptor level in (Ga,Mn)As is well established to be located at 110 meV above the valence-band (VB) edge, which is much higher than the levels of shallow acceptors, *e.g.* Zn. The point of the debate is whether insulator to metal transition occurs in the merged valence and impurity bands (VB model) or in the still detached but well resolved impurity band (IB model) [6, 12]. According to the VB model the valence band edge should merge with the Mn impurity band which was indirectly illustrated in various experiments [13-17] and justified by theoretical calculations [11]. By adapting the tight-binding Anderson approach and full-potential local-density approximation +$U$ calculations, J. Mašek *et al.* have excluded the validity of the IB model in ferromagnetic (Ga,Mn)As at all [11]. On the other hand, many results obtained by different spectroscopic studies, *e.g.* ellipsometric studies [18], infrared conductivity measurements [19], and more recently, magnetic circular dichroism [6] and resonant tunnelling spectroscopy [20] strongly support the IB model. Nowadays, low-temperature molecular beam epitaxy (LT-MBE) is the most commonly used method to fabricate ferromagnetic (Ga,Mn)As samples. As an alternative, ion implantation followed by nanosecond pulsed laser annealing (PLA) was utilised to produce ferromagnetic DMS as well [21, 22]. These two methods allow Mn doping well beyond the equilibrium solubility limit, at the same time avoiding phase segregation and maintaining a relatively high degree of uniformity and crystal quality.

In this paper, we utilize ion beam implantation and millisecond range flash lamp annealing (FLA) to fabricate (Ga,Mn)As samples with a Mn content varying from 0.02 to 0.6%. During millisecond range FLA the top layer of the implanted GaAs sample is heated to a temperature slightly lower than the melting point of GaAs, which significantly enhances the solubility of Mn in GaAs. Simultaneously, due to the millisecond range solid-phase epitaxial regrowth of implanted GaAs, the diffusion of Mn and the formation of As related defects are significantly suppressed. Consequently, high-quality (Ga,Mn)As layers can be formed which exhibit room temperature photoluminescence (RTPL) and well-defined critical points in ellipsometry spectra. The crystalline quality of Mn implanted and annealed samples was confirmed by X-ray diffraction spectroscopy and high resolution TEM investigation. These samples allow characterising the alterations of the GaAs VB upon Mn doping. Based on the RTPL and ellipsometry data, the upward movement of the VB in (Ga,Mn)As with increasing Mn concentration is confirmed, supporting the VB model proposed by Dietl [3]. Millisecond range FLA represents a compromise between the high doping concentration and the crystalline quality.

## 2. Experimental part

Semi-insulating, (001) oriented GaAs wafers were implanted with 300 keV Mn ions at fluences corresponding to peak concentrations of 0.02, 0.04, 0.3 and 0.6 at.%. The projected range, $R_p$, for Mn is around 200 nm below the surface. After ion implantation the samples were flash-lamp annealed for 3 ms with an energy density of 40 $Jcm^{-2}$. The maximum temperature obtained during FLA was in the range of 1100 – 1150 $^o$C. The optical spectrum of the Xe lamps in our FLA system has a maximum emission at a wavelength of about 500 nm with maximum energy density of about 120 $Jcm^{-2}$. The pulse duration is controlled by the parameters of the LC-network. The detailed description of the model used to simulate the temperature distribution in flashed samples and description of our FLA system can be found in ref. 23 and 24, respectively. Figure 1 shows the temperature distribution as a function of time in GaAs wafer during 3 ms flash lamp annealing with an energy density of 40 Jcm$^{-2}$ at different depth. Black curve (0 nm) shows the temperature at the surface while the curve 400 µm shows the change of temperature at the back side of the annealed wafer (see Fig. 1). Here the cooling is determined only by radiative loses and heat conduction inside the wafer.

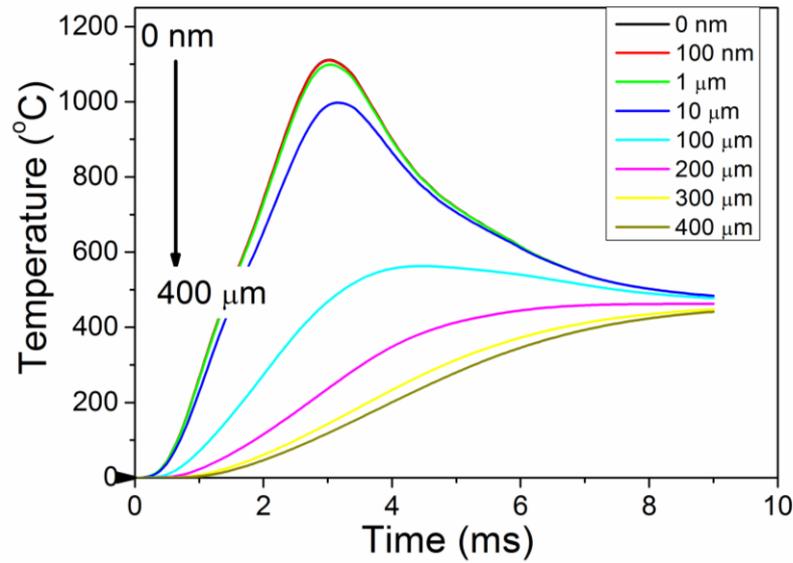

Figure 1. (Color online) Simulated temperature distribution at different depths in a GaAs wafer during 3 ms flash lamp annealing with an energy density of 40 Jcm$^{-2}$.

The phonon spectra were determined by micro-Raman spectroscopy in backscattering geometry in the range of 100 to 700 cm$^{-1}$ using a 532 nm Nd:YAG laser with a liquid-nitrogen cooled charge coupled device camera. The PL measurements were performed under a blue (405 nm) or green (532 nm) laser illumination with a maximum excitation power up to 150 mW. The PL spectra were recorded at a temperature ranging from 15 to 300 K using a Jobin Yvon Triax 550 monochromator and a cooled InGaAs detector. For the ellipsometric measurements a rotating analyser ellipsometer VASE system (J. A.Woollam Co., Inc., USA) operating at an energy range of 1 to 4 eV was used. The conductivity type and carrier concentration were determined by Hall effect measurements. X-ray absorption spectroscopy (XAS) measurements were performed at the beamline UE46/PGM-1 at BESSY II (Helmholtz-Zentrum Berlin). A magnetic field up to 10 kOe was applied parallel to the photon helicity and perpendicular to the surface. The DC magnetization measurements were carried out as a function of temperature using a superconducting quantum interference device magnetometer (SQUID-VSM). The magnetic field was applied in plane parallel to the sample surface. Transmission electron microscopy (TEM) investigations were performed in cross-sectional geometry by means of an image-corrected Titan 80‑300 microscope (FEI) operating at an accelerating voltage of 300 kV. X-ray diffraction (XRD) was performed by a Empyrean Panalytical diffractometer with a Cu-target source. The setup is equipped with a Göbel mirror and a asymmetric Monochromator to enhance the brilliance and monochromaticy. The distribution of Mn atoms in GaAs matrix was investigated by means of secondary ion mass spectrometry (SIMS).

## 3. Results and discussion

### 3a. Optical properties of Mn doped (Ga,Mn)As

Assuming the VB model to be valid, the peak position of the near-band edge emission of the Mn-implanted layer should be strongly affected by the Mn doping and exhibit a red-shift with increasing Mn content. Figure 2a shows the RTPL spectra taken from samples containing different Mn concentration during green laser (532 nm) excitation. The PL spectra consist of two bands at 875 nm and at 910 ±15 nm. The projected Mn distribution in the as-implanted samples is peaking around 200 nm below the surface while the penetration depth of the green laser light in GaAs is more than 1.5 μm [25]. Consequently, the PL spectra taken from our samples are superimposed PL emissions from both the undoped GaAs substrate and the (Ga,Mn)As layer. The PL peak at 875 nm originates from the near band emission of the GaAs substrate and is independent of the Mn concentration, while the second PL band exhibits a red shift with increasing Mn concentration.

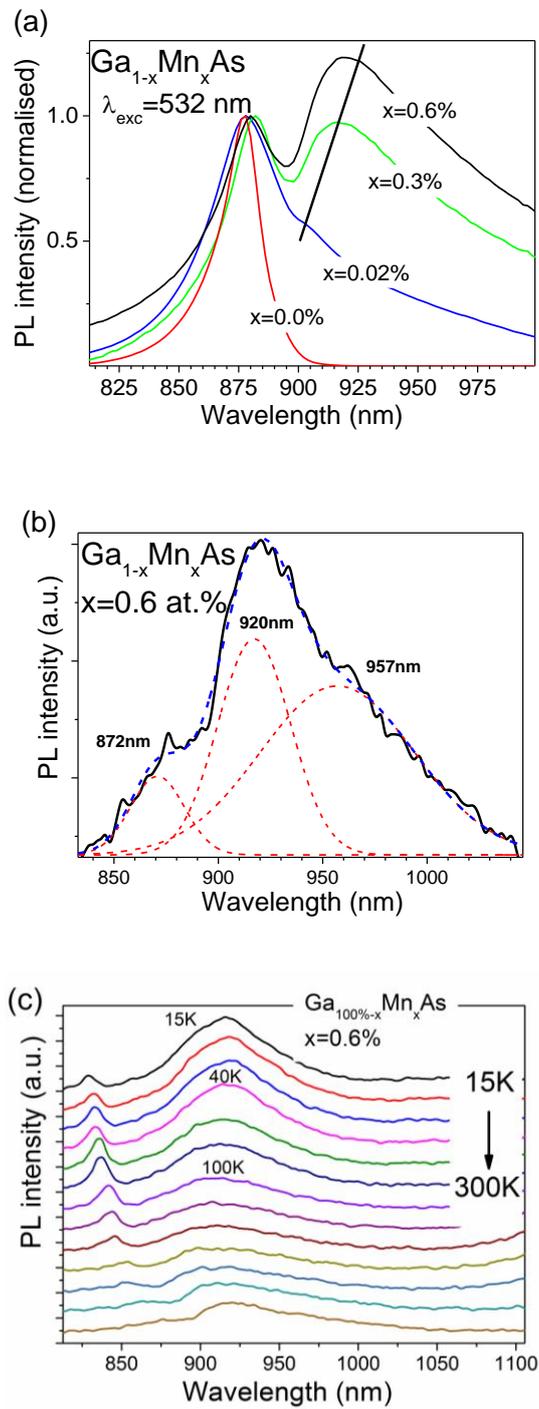

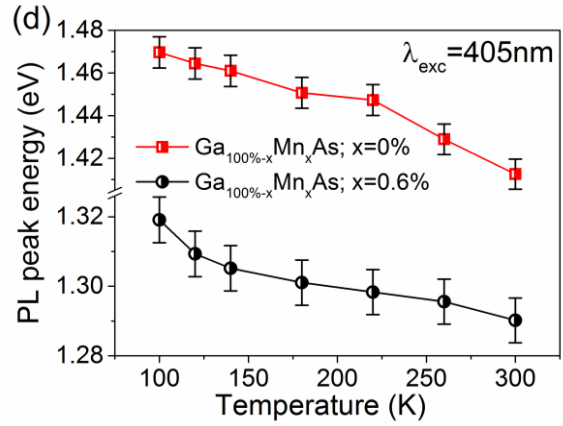

Figure 2. (Color online) RTPL obtained from $Ga_{100\%-x}Mn_xAs$ with x=0.0, 0.02, 0.3 and 0.6 at.% after excitation with green laser light (532 nm) (a). The RTPL with Gaussian deconvolution of PL spectrum and TDPL spectra obtained from samples containing 0.6 at.% of Mn after blue laser excitation (405 nm) (b) and (c), respectively and the PL energy peak position of the near band emission as a function of temperature after laser excitation at 405 nm (d).

The PL emission at around 910 ±15 nm is assigned to the near band emission from the $Ga_{100\%-x}Mn_xAs$ layer due to the upward shift of the valence band edge with increasing Mn concentration. In order to suppress the influence of the substrate on the PL data, a blue laser (405 nm) was used for excitation (see Fig. 2b). In case of the 405 nm laser the penetration depth is comparable to the projected range, $R_p$, and the emission from the highly-doped region dominates the PL spectrum. Similar to green laser excitation, by increasing the Mn doping a significant red-shift of the PL peak position is observed (see Fig. 2b). The same phenomenon takes place in p-type GaAs heavily doped with shallow dopants e.g. Zn or Be which fully supports our assumption about band gap narrowing in $Ga_{100\%-x}Mn_xAs$ alloys [26]. The PL spectrum obtained from the $Ga_{100\%-x}Mn_xAs$ layer containing 0.6 at.% of Mn was fitted with three Gaussian functions. Those three bands can be distinguished (i) at about 872 nm observed from the GaAs substrate, (ii) at 921 nm due to near band emission from lightly doped region in (Ga,Mn)As and (iii) band at 957 nm (Ga,Mn)As alloy where the VB is merged with the IB. In the case of an ion beam implanted sample, before annealing, the implanted elements exhibit a Gaussian depth distribution (see green curve in Fig. 7a). It means that the maximum concentration of the implanted element is usually obtained at $R_p$, while above and below $R_p$ the impurity concentration tends towards zero. After FLA, the Mn atoms are redistributed into two regions: highly doped region starting from the sample surface up to 200 nm and lightly doped region below 200 nm (see red curve in Fig. 7a). Such redistribution of Mn atoms indicates the existence of two (Ga,Mn)As layers with the lightly and heavily affected band gaps of GaAs, respectively. This explains the co-existence of two bands in the PL spectrum obtained during 405 nm laser excitation, when only the implanted layer should be excited. The emission at 921 nm originates from the lightly doped region while the 957 nm emission is from the top highly Mn doped layer (see Fig. 2b) which directly proves the VB scenario in the Mn doped GaAs.

We have also investigated the photoluminescence obtained from Mn doped GaAs at temperatures from 15 K to 300 K (see Fig. 2c). The PL spectra taken below 100 K consist of the near band emission, conduction-band-to-Mn acceptor transitions and defect related luminescence. For example, intrinsic defects formed during the ion irradiation, like the optically active arsenic and gallium vacancies dominate the low temperature PL spectra [27, 28]. The energy levels corresponding to those defects are located between 30 and 160 meV below the conduction band of GaAs and the low temperature PL emission peaks are located between 840 and 960 nm. Above 100 K most of defect related optical emissions in GaAs are thermally quenched and in the near infrared range only the near band emission is visible [29]. Therefore, the interpretation of the PL data measured from Mn implanted GaAs below 80 K is ambiguous. The low temperature PL emission from Mn implanted GaAs were investigated in details by Yu and Park [30]. They have shown that the PL peak located at about 95±15 meV below the band gap of GaAs corresponds to the conduction-band-to-Mn acceptor transitions and the emission intensity is rapidly quenched above 50 K. Figure 2d shows the PL energy peak position of the near band emission as a function of temperature for GaAs substrate (red squares) and $Ga_{100\%-x}Mn_xAs$ layer with $x$=0.6 at.% (black circles) after blue laser excitation. The energy peak position, obtained from the Mn doped layer after Gaussian deconvolution of the PL spectra taken at different temperatures shows a red-shift with increasing temperature which is typical for the band-to-band transition. Luminescence taken from the undoped GaAs substrate shows the same behaviour under the same measurement conditions. Both the concentration dependence and the red-shift of the peak position with temperature establish direct evidence that Mn strongly affects the VB in the (Ga,Mn)As alloy, even for low Mn concentrations (<1 at.%), which is in contrast to the IB model where the Mn doping has little influence on the VB edge [31]. An interesting study has been recently reported by Yastrubchak *et al.*, who investigated LT-MBE $Ga_{100\%-x}Mn_xAs$ samples with a nominal concentration of Mn up to 1.2 at.% by modulation photoreflectance spectroscopy [16]. They have shown that already 0.001 at.% of Mn influences the electronic structure of GaAs and that the IB merges with the VB for a Mn concentration higher than 0.005 at.%, although the $Ga_{100\%-x}Mn_xAs$ layer shows n-type conductivity due to the high-concentration of arsenic antisites $As_{Ga}$.

Spectroscopic ellipsometry represents another method to investigate the band structure of semiconductors. However, due to the low crystalline quality of the heavily Mn doped GaAs layer grown by LT-MBE, the signatures of the main critical points in $Ga_{100\%-x}Mn_xAs$ are often smeared out [18]. We show here that the ellipsometry data has been significantly improved in flash lamp annealed Mn implanted GaAs.

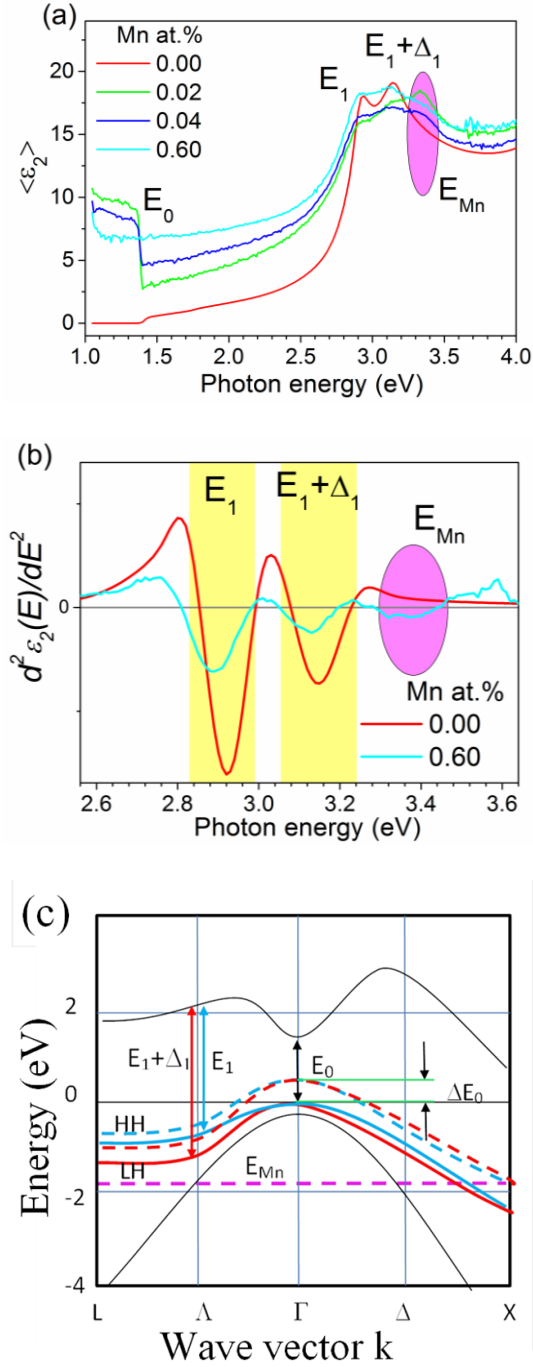

Figure 3. (Color online) Imaginary part of the dielectric function measured at 80 deg angle of incidence taken from $Ga_{100\%-x}Mn_xAs$ with 0<x<0.6 at.% along GaAs [$\bar{1}$10] direction (a) and the high-energy part of the second derivative ($d^2\varepsilon_2(E)/dE^2$) spectra (b). (c) A schematic drawing of the (Ga,Mn)As band structure with critical point transitions. The valence bands are indicated as HH (blue) for heavy-holes and LH (red) for light holes. Solid and dashed lines correspond to bands in virgin GaAs and Mn-doped GaAs, respectively. The dispersion less $E_{Mn}$ acceptor level (occupied spin-down Mn *d*-shell level) is also shown (dashed-magenta line in (c)).

Figure 3a shows the imaginary part of the pseudodielectric function measured at an incident angle of 80 degree taken from a virgin GaAs wafer and from a (Ga,Mn)As sample containing up to 0.6% of Mn. The measurements were performed along GaAs [$\bar{1}$10]. Significant changes in the ellipsometric spectra are due to the enhancement of the interband transitions in the vicinity of critical points in the Brillouin zone (see Fig. 3c). In contrast to LT-MBE prepared (Ga,Mn)As [18] or ion implanted GaAs subsequently annealed by nanoseconds laser [32], in the FLA treated sample all critical points are clearly resolved independent of the incident angle confirming the high-quality of the (Ga,Mn)As layer.

As shown in Fig. 3a, in the low-energy part (E<2.5 eV) the photon absorption increases significantly with increasing *x*. This process is enhanced for the measurements taken at the incident angle of 80 degree along GaAs [$\bar{1}$10] (above the Brewster angle ($\theta_B$)), when the signal is detected mainly from the Mn implanted layer. The increase of optical absorption in the range of 1.4 – 2.5 eV must be caused by the modification of the band edge and tail. The high-energy part of the dielectric function consists of two peaks at around 2.9 and 3.1 eV corresponding to $E_1$ and $E_1+\Delta_1$ transitions near the $\Lambda$ point, respectively. Figure 3b shows the second derivative of imaginary part of the dielectric function measured at the incident angle of 80 degree taken from the virgin and highly-doped sample along the GaAs [$\bar{1}$10] direction. In contrast to the data of Burch *et al.* [18], our ellipsometric results clearly show a red shift of the critical points with increasing Mn concentration. In the case of the $E_1$ and $E_1+\Delta_1$ transitions the shift is up to 49 meV and 36 meV for 0.6 at.% of Mn, respectively, related to the non-degeneracy between heavy-holes (HHs) and light-holes (LHs) at the $\Lambda$ point. The non-degeneracy between HHs and LHs is also observed for the samples with 0.02 at.% (not shown) and 0.04 at.% Mn (Fig. 3c), which have lower hole concentrations of around $4\times10^{18}$ and $8\times10^{18}$ cm$^{-3}$, respectively, as confirmed by Hall measurements and Raman spectroscopy (see Fig. 4). We can schematically draw the band structure for Mn doped GaAs as shown in Fig. 3c. The change of the critical-point position with Mn concentration directly correlates with the upward shift of the valence band near the $\Lambda$ point. To this end, the red shift of critical points correlates well with PL data and fully supports the VB model proposed by Dietl [3]. It is worth to note that for Mn doped samples, an additional peak at around 3.4 eV can be distinguished (see Fig. 3a and b). Since it appears only in the Mn doped samples, we assign it to the Mn related transition in the (Ga,Mn)As layer ($E_{Mn}$), *i.e.* the occupied spin-down Mn *d*-shell level located deep in the valence band (see Fig. 3c) [33].

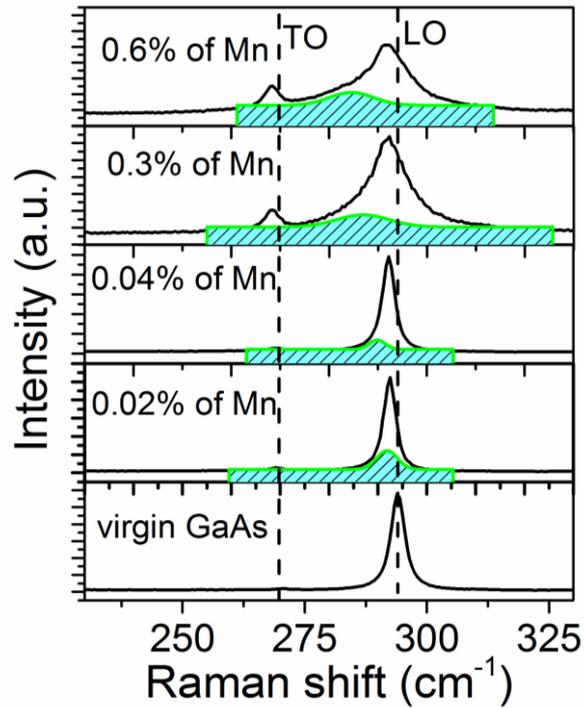

Figure 4. (Color online) Micro-Raman spectra obtained from (Ga,Mn)As alloys with different Mn concentration. The coupled LO-phonon plasmon (CLOPM) mode obtained by Gaussian deconvolution of the Raman spectra is shown as the green sub-spectra. The mode is shifted to low frequency range with increasing Mn concentration, i.e. hole concentration.

The conductivity type and structural properties of Mn doped GaAs were investigated by means of micro-Raman spectroscopy (see Fig. 4). The Raman spectra taken from (Ga,Mn)As show the TO and LO phonon modes whose positions exhibit a red shift by 2 cm$^{-1}$ compared to the virgin GaAs. Moreover, the LO-like phonon mode exhibits a strong asymmetry on the left side due to a coupling with the plasmon mode (see Fig. 4). This is typical for p-type GaAs with a carrier concentration higher than $10^{18}$ cm$^{-3}$ [34]. The coupled LO-phonon plasmon mode (CLOPM) shifts to lower frequencies with increasing Mn concentration. Using Gaussian deconvolution of the Raman spectrum taken from the sample containing 0.6 at.% of Mn, the CLOPM is estimated to be at 282 cm$^{-1}$ which corresponds to a hole concentration of ~$3\times10^{19}$ cm$^{-3}$ [35]. The hole concentration and conductivity type of (Ga,Mn)As layer were confirmed by Hall effect measurements but the highest hole concentration measured by Hall effect was about $8.6\times10^{18}$ cm$^{-3}$ with the hall mobility about 77 cm$^2$/(V·s). The temperature dependent resistivity obtained from Mn doped GaAs samples shows typical semiconducting behaviour. Nevertheless our sample with the Mn concentration of 0.6 at.% is still in the insulating side. However, the hole concentration and the resistivity are in the same order of magnitude compared to a sample grown by MBE with Mn concentration of 1.5% [36]. This also proves that the activation fraction of Mn is high and the compensation is low in our sample. According to Hall effect measurements we found that after ion implantation and flash lamp annealing a Mn doping above 0.02 at.% is sufficient to compensate any type of unintentionally present donors in GaAs.

**3b. Magnetic properties of GaMnAs alloys.**

To prove that there is no MnAs formed during flash lamp annealing, we have performed X-ray absorption spectroscopy (XAS) utilizing circularly polarized light and magnetization measurement on the GaMnAs sample containing 0.6 at.% of Mn in which the VB starts to merge with IB. The XAS directly probes the electronic structure of the polarized Mn 3*d* band and it is only sensitive to the surface region (up to a few nm) in total electron yield mode. As illustrated in Fig. 5a, the XAS spectra show clearly multiple structures, which are completely different from the spectrum for MnAs with broad, featureless peaks [37]. Our XAS spectrum resembles the typical feature of localized Mn $3d^5$ [38]. The slightly more pronounced multiple structures compared to Ref. 33 are very probably due to the residual oxidized Mn at the sample surface even thought we did HCl etching right before XAS measurements to removed oxide layer. However, the sizeable X-ray magnetic circular dichroism (XMCD) at the Mn $L_{2,3}$ edges, *i.e.* the difference between XAS when changing the helicity of the incoming photons, clearly supports the formation of paramagnetic GaMnAs (see Fig. 5b) [38].

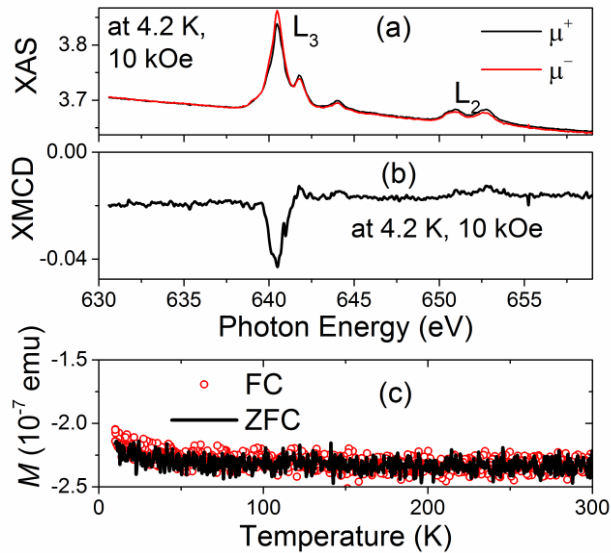

Figure 5. (Color online) Mn $L_{3,2}$ total electron yield (a) XAS for magnetization and helicity parallel ($\mu^+$) and antiparallel ($\mu^-$) and (b) XMCD ($\mu^+$-$\mu^-$) for (Ga,Mn)As containing 0.6 at.% of Mn measured at around 4.2 K under an external magnetic field applied parallel to the surface. (c) temperature dependent magnetic moment (*M*) measured under 50 Oe using ZFC/FC protocol. The negative sign of the ZFC/FC magnetization is due to diamagnetic substrate.

Moreover, the shape of the XMCD spectrum is very similar to that of (Ga,Mn)As prepared by MBE [38, 39]. The small dichroism ratio is also due to the surface oxidization and due to the paramagnetic nature. While the XAS in total electron mode is only surface sensitive, we also performed bulk sensitive magnetometry measurements. Fig. 5c shows the zero field cooling and field cooling (ZFC/FC) magnetic moment measured under a field of 50 Oe following the protocol in Ref. [40]. If there were MnAs nanocrystals embedded in GaAs, one should observe irreversibility in ZFC/FC magnetization curves. However, our sample does not reveal any difference between FC and ZFC curves, excluding the formation of MnAs nanocrystals.

**3c. Structural investigation of (Ga,Mn)As alloys**

Both RTPL and micro-Raman spectroscopy data suggest the existence of a high-quality (Ga,Mn)As layer. Simultaneously, SQUID investigations exclude secondary phase formation of ferromagnetic nanoparticles like MnAs. For further investigations of the influence of the millisecond-range FLA on the crystallinity and lattice variations, high-resolution θ-2θ XRD scans around GaAs (004) were performed for the as-implanted and FLA-treated (Ga,Mn)As samples. Figure 6 shows the experimental and simulation data obtained from samples containing 0.3 at.% (a) and 0.6 at.% (b) of Mn in the as-implanted state and after annealing, respectively. After Mn implantation the top GaAs layer becomes strained along the surface normal (parallel to [001] direction) due to damages introduced into the sample during the ion bombardment process. Many of the Ga and As atoms are displaced from their initial positions leading to the formation of an uniaxial strained layer. The shift of the lattice expansion peak towards lower scattering angles describes that lattice expansion along [001]. For the sample with 0.3 at.% implanted (Fig.6a) thickness fringes after implantation can be clearly resolved where as for the as implanted with 0.6 at.% sample (Fig. 6b) thickness fringes are almost suppressed indicating possibly the begin of plastic deformation. The strained layers are almost disappeared after flash lamp annealing. The 0.3 at.% sample shows only a small asymmetry on the left side of the main peak. Such a phenomenon is well known in (Ga,Mn)As alloys and is assigned to a change of the lattice parameter of the GaAs host due to the incorporation of Mn on Ga sites during annealing [41]. Interestingly, after annealing the shoulder on the right side of the main diffraction peak remains.

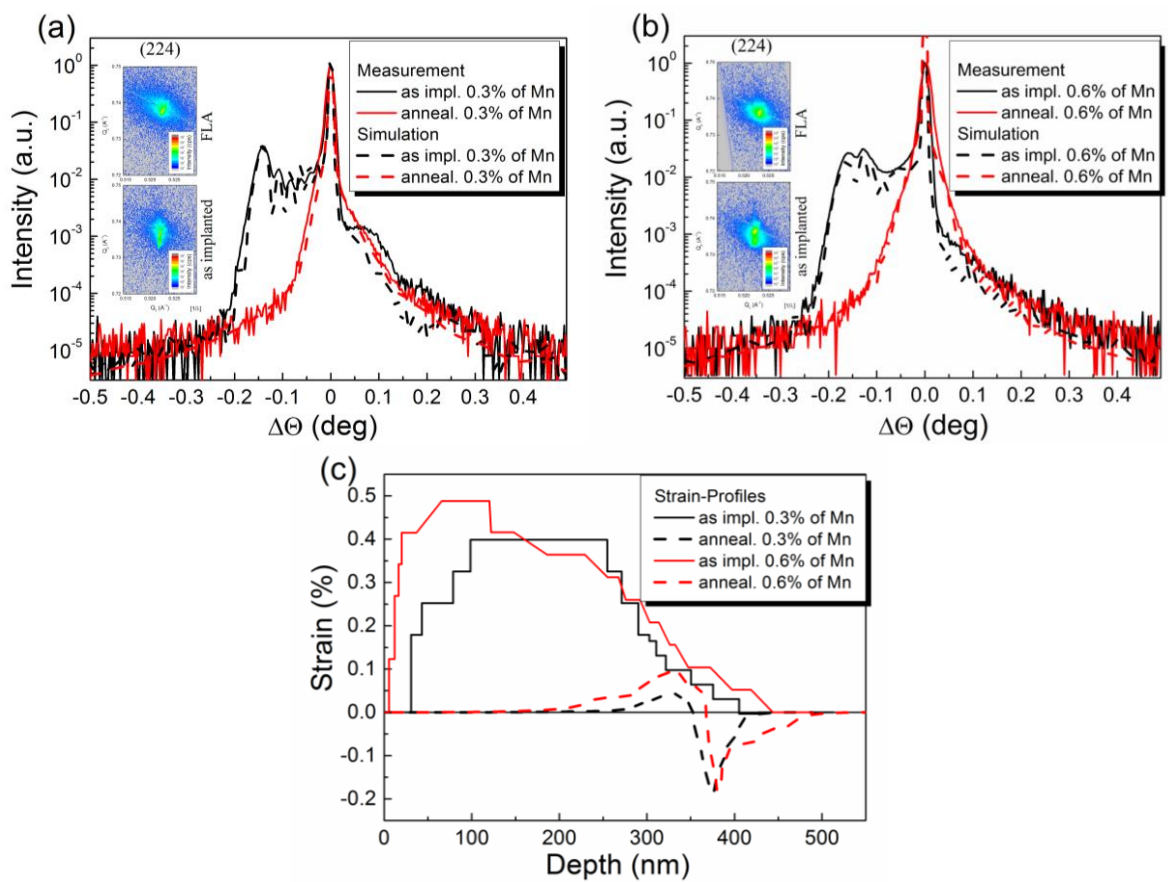

Figure 6. (Color online) XRD θ–2θ scans obtained from Mn-implanted GaAs wafers before and after annealing. Samples were flash lamp annealed for 3 ms with an energy density of 40 Jcm$^{-2}$. The experimental and simulation data obtained from samples containing 0.3 at.% and 0.6 at.% of Mn are shown in (a) and (b), respectively. (c) shows strain profiles by simulating the XRD data in as-implanted (solid line) and annealed (dotted line) (Ga,Mn)As samples for two different Mn concentration 0.3 at.% and 0.6 at.%. The insets in (a) and (b) show reciprocal space mapping (RSM) of (224) diffraction peak from (Ga,Mn)As samples taken before and after annealing.

The intensity of this diffraction signal decreased after annealing but it is still detectable demonstrating the annealing effect, i.e. the strong reduction of defects upon FLA. The 0.6at.% sample shows a broader diffuse background after FLA indicating again a larger number of defects introduced after implantation which was not completely removed after FLA.

Reciprocal space maps around the (224) reflection of GaAs are shown in the insets (Fig. 6a and 6c). These maps illustrate again that implantation leads to the formation of a uniaxial strained layer and that annealing lead to an almost complete recovery.

To quantify this effect we have simulated the strain distribution of the 0.3 at.% and 0.6 at.% samples (Fig. 6c.). In the as-implanted stage the strained layer is formed within the projected ion range, up to 300 nm from the sample surface. Whereas the depth of the strained layer did not depend on the fluence we have defined consequently a slightly higher (>15%) uniaxial strain for the 0.6 at.% sample. During FLA the damaged layers recover almost to the initial single crystalline phase; the strain is released and at the same time implanted Mn is at least partially introduced into the lattice sites. Moreover, the annealed samples contain a thin strained layer with local compressive and tensile strains (positive and negative signal in Fig. 6c) about 350 nm below the surface. The strain accumulated there is due to the end-of-range defects typical for ion implantation [42]. Since those defects are located below the Mn doped layer, they have no influence on the optical and magnetic properties presented in this paper.

The formation of a single-crystalline almost defect-free (Ga,Mn)As layer by ion beam implantation and ms-range FLA was confirmed by TEM investigation. Figure 7a shows a cross-sectional bright-field TEM micrograph obtained in [110] zone axis geometry for the annealed (Ga,Mn)As sample containing 0.6 at.% of Mn. As can be seen, the originally Mn-implanted sub-surface region is homogeneous in crystal structure without any signs for the formation of secondary phases. For detailed analysis, figure 7b shows a high-resolution TEM (HRTEM) image taken about 150 nm below the sample surface. According to this data, the implanted and FLA-treated (Ga,Mn)As sample is of single-crystalline nature having the same crystal orientation as the GaAs substrate. Below the implantation region, i.e. about 250 nm from the sample surface, only the end-of-range defects are visible [43]. Such defects are always present in implanted and subsequently annealed semiconductors. The existence of end-of-range defects visible in fig. 7a is in accordance with the presented XRD data showing up there as a strained layer (see Fig. 6c). Since those defects are located beneath the Mn-doped layer, they do not have any influence on the presented optical and magnetic data. The two curves superimposed artificially to the TEM micrograph in Fig. 7a show the Mn distribution in GaAs obtained by SIMS from the as-implanted sample (green curve) and after FLA treatment (red curve). It should be mentioned that the Mn surface peak visible for the annealed sample is a measuring artefact resulting from nonhomogeneous sputtering of close to the surface region of implanted and annealed GaAs. Having in mind that the maximum temperature obtained in the sample is in the range of 1100 $^o$C, it should not be surprising to have some Mn redistribution during annealing.

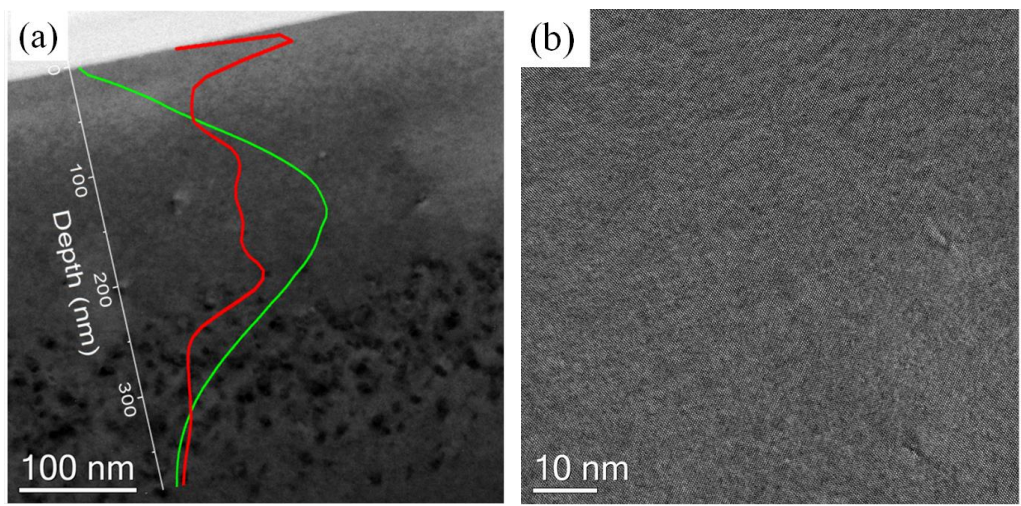

Figure 7. (Color online) (a) Cross-sectional bright-field TEM micrograph taken from (Ga,Mn)As containing 0.6 at.% of Mn after FLA annealing for 3 ms with an energy density of 40 Jcm$^{-2}$. The superimposed green and red curves are the Mn distributions in the as-implanted

sample and after FLA treatment, respectively, which were obtained by SIMS. Part (b) shows a HRTEM image taken from an area about 150 nm below the sample surface.

Taking into account the diffusion coefficient of Mn in GaAs ($5\times10^{-10}$ cm$^2$/s at 1100 $^o$C from [44]) and the temperature distribution in our sample presented in Fig. 1, the single Mn atom cannot diffuse more than 6 nm during FLA. Moreover, in the sample containing 0.6 at.% of Mn, the average distance between Mn atoms in the as-implanted stage is about 4 nm which excludes the possibility of ferromagnetic cluster formation during the heat treatments performed in this study. The presented microstructural data of Mn-implanted and FLA-treated GaAs fully support the interpretation of our optical and magnetic data shown above.

**Conclusions**

Based on room-temperature photoluminescence and spectral ellipsometry data, the band gap narrowing in (Ga,Mn)As alloy can be explained with the VB model proposed by T. Dietl [3]. We have shown that even Mn concentrations as low as 0.02 at.% are sufficient to influence the VB edge. For Mn concentrations above 0.6 at.%, the VB merges with the IB. Microstructural investigations exclude secondary phase formation in Mn-implanted and FLA-treated GaAs.

**Acknowledgement**

This work was financially supported by the Helmholtz-Gemeinschaft Deutscher Forschungszentren (HGF-VH-NG-713) and by DFG (SCHM1663/4-1). We would like to thank the Ion Beam Center at Helmholtz-Zentrum Dresden-Rossendorf for the ion implantation.

**Figure legends**

Figure 1. (Color online) Simulated temperature distribution at different depth in GaAs wafer during 3 ms flash lamp annealing with an energy density of 40 Jcm$^{-2}$.

Figure 2. (Color online) RTPL obtained from Ga$_{100\%-x}$Mn$_x$As with x=0.0, 0.02, 0.3 and 0.6 at.% after excitation with green laser light (532 nm) (a). The RTPL with Gaussian deconvolution of PL spectrum and TDPL spectra obtained from samples containing 0.6 at.% of Mn after blue laser excitation (405 nm) (b) and (c), respectively and the PL energy peak position of the near band emission as a function of temperature after laser excitation at 405 nm (d).

Figure 3. (Color online) Imaginary part of the dielectric function measured at 80 deg angle of incidence taken from Ga$_{100\%-x}$Mn$_x$As with 0<x<0.6 at.% along GaAs [$\bar{1}$10] direction (a) and the high-energy part of the second derivative ($d^2\varepsilon_2(E)/dE^2$) spectra (b). (c) A schematic drawing of the (Ga,Mn)As band structure with critical point transitions. The valence bands are indicated as HH (blue) for heavy-holes and LH (red) for light holes. Solid and dashed lines correspond to bands in virgin GaAs and Mn-doped GaAs, respectively. The dispersion less E$_{Mn}$ acceptor level (occupied spin-down Mn *d*-shell level) is also shown (dashed-magenta line in (c)).

Figure 4. (Color online) Micro-Raman spectra obtained from GaMnAs alloys with different Mn concentration. The coupled LO-phonon plasmon (CLOPM) mode obtained by Gaussian deconvolution of the Raman spectra is shown as the green sub-spectra. The mode is shifted to low frequency range with increasing Mn concentration, i.e. hole concentration.

Figure 5. (Color online) Mn $L_{3,2}$ total electron yield (a) XAS for magnetization and helicity parallel ($\mu^+$) and antiparallel ($\mu^-$) and (b) XMCD ($\mu^+$-$\mu^-$) for GaMnAs containing 0.6% of Mn measured at around 4.2 K under an external magnetic field applied parallel to the surface. (c) temperature dependent magnetic moment (*M*) measured under 50 Oe using ZFC/FC protocol. The negative sign of the ZFC/FC magnetization is due to diamagnetic substrate.

Figure 6. (Color online) XRD pattern obtained from Mn implanted GaAs wafers before and after annealing. Samples were flash lamp annealed for 3 ms with an energy density of 40 Jcm$^{-2}$. (a) shows the measured data, (b) simulation and (c) shows strain profiles in as-implanted and annealed GaMnAs samples for two different Mn concentration 0.3% and 0.6%. Solid lines show the strain profile calculated from XRD data, while the dashed lines are obtained by simulation.

Figure 7. (Color online) (a) Cross-sectional bright-field TEM micrograph taken from (Ga,Mn)As containing 0.6 at.% of Mn after FLA annealing for 3 ms with an energy density of 40 Jcm$^{-2}$. The superimposed green and red curves are the Mn distributions in the as-implanted sample and after FLA treatment, respectively, which were obtained by SIMS. Part (b) shows a HRTEM image taken from an area about 150 nm below the sample surface.